# The Feasibility of Algorithmic Detection and Decentralised Moderation for Protecting Women from Online Abuse

## November 2022

Sarah Barrington (MIMS23)

University of California, Berkeley

A Work In Progress Methodology and Pre-Print

## I. ABSTRACT

Online abuse is becoming an increasingly prevalent issue in modern day society, with 41% of Americans having experienced online harassment in some capacity in 2021[1]. People who identify as women, in particular, can be subjected to a wide range of abusive behavior online, with gender-specific experiences cited broadly in recent literature across fields such as blogging[2], politics[3] and journalism[4].

In response to this rise in abusive content, platforms have been found to largely employ '*individualistic moderation*' approaches, aiming to protect users from harmful content through the screening and management of singular interactions or accounts[5]. Yet, previous work performed by the author of this paper has shown that in the cases of women in particular, these approaches can often be ineffective; failing to protect users from *multi-dimensional abuse* spanning prolonged time periods, different platforms and varying interaction types.

In recognition of its increasing complexity, platforms are beginning to outsource content moderation to users in a new and *decentralized* approach. The goal of this research is to examine the feasibility of using multidimensional abuse indicators in a Twitter-based moderation algorithm aiming to protect women from female-targeted online abuse. This research outlines three indicators of multidimensional abuse, explores how these indicators can be extracted as features from Twitter data, and proposes a technical framework for deploying an end-to-end moderation algorithm using these features.

## II. INTRODUCTION

The centralized moderation approaches of mainstream social media platforms are becoming increasingly scrutinized in both legal and sociological literature, and are often linked to problematic phenomena such as 'content cartels'[6], the amplification of misinformation and political polarization[7]. Current algorithms that curate and moderate content online are typically considered as 'one size fits all', and as such, can exhibit technological vulnerabilities that can be exploited, and in particular, be perceived to alter the online experiences of marginalized groups[8].

A prior, independent qualitative research study was performed by the authors to explore this specific moderation context within the marginalized community of females who experience online abuse[9].

The objective of this study was to explore the perception of threat in a range of online abuse situations through the contexts, themes and experience of five self-identifying women. Three key findings emerged that defined the context of female-targeted online abuse in a novel way:

- Abuse manifests as an ecosystem comprising of three components - a mixture of dichotomous roles, motivations (in which the majority were objectification-related) and complex 'arenas' that can span across platforms, time and the digital-physical world boundary.

- Online threats become 'real' threats through the rise of two asymmetries- volumetric, in which interactions on social media platforms were observed to 'spiral out of [the target's] control', and informational, in which the abuser(s) are perceived to possess more identifying information about the target than the target had about the abuser.
- The above two phenomena are currently overlooked due to significant limitations in platform-bound control & response strategies- platforms commonly assume a uni-directional relationship between the abuser and target, with little longitudinal context of how the abuse or the relationship evolves- meaning that women often had to combine multiple interventions in order to develop their own personal safety strategies.

Each of these technological phenomena were observed to contribute towards the emergence of power imbalances, that were further exacerbated by underlying patriarchal constructs which played a part in defining the complex, nuanced relationship between the roles of abuser and target.

Despite platforms offering multiple tools for promoting online safety, and participants sharing awareness of these, the results from the foundational study suggested that these tools exist on an individualistic level- by individual account (blocking, reporting) or interaction (flagging, filtering, deleting). These findings show how online abuse can be a product of compounded multidimensional mechanisms and power imbalances, that can evolve through time and lead to direct, real-world impacts. As such, it may be ineffective to define abuse at the individual interaction and account levels.

The concept of 'individualistic moderation' can be observed as a somewhat nascent ideology in wider literature, addressed briefly and conceptually by moderation experts such as Prof. Tarleton Gillespie of the Cornell Communications Department (*Custodians of the Internet*[10]). Generally, the few references to solutions are brief and often idealistic; aiming to provoke thought and discussion. For example, Political Scientist Dr. Francis Fukuyama, at the Stanford Center on Democracy, Development and the Rule of Law, makes a call for *middleware* innovation; in which decentralized technologies allow users and communities to regain control over their content moderation. However, there is little detail on the specifics of how specific technological solutions might work, and how they might tangibly help affected communities online.

By contrast, multidimensional indicators must account for the interaction history and dynamic between the abuser and target. The above findings can be summarized into the following indicators:
- **Volumetric asymmetry** - in which interactions on social media platforms were observed to 'spiral out of the target's control'
- **Informational asymmetry** - in which the abuser(s) are perceived to possess more identifying information about the target than the target had about the abuser
- **Longitudinal relationship** - involving the evolution of multiple historic interactions over a prolonged time period (which was user-specific, sometimes noted as a year or longer).

This research explores how these indicators can be extracted as features from Twitter data, and proposes a technical framework for deploying an end-to-end moderation algorithm using these features.

The novel contributions of this paper are as follows:
- Moderation algorithm based upon the encoding of indicators to data features and API calls
- Suggests platform actions that align to the indicators
- Proposes an architecture for moderation end to end using Twitter
- Summary of the limitations and feasibility of deploying this

## III. METHODS

The aim of this work is to evaluate the feasibility of decentralized (non-platform) moderation to protect women from female-targeted online abuse. This involved three separate parts: gathering data to train a python-based moderation algorithm, building an end-to-end pipeline that could be used to detect any least one of the multidimensional indicators, and perform actions through a social media API. Twitter has been selected as the initial prototype platform for this work due to its extensive research access and broad array of API end points.

### Data Acquisition

Having obtained full research access (tweet cap of 10million tweets per month), a short list of abusive content is required in order to provide training data that exhibited problematic content (content that would likely be moderated by a female user).

To facilitate this process, an initial list of abuse-centric tweets is imported from the *twitter-cyberbullying* dataset curated by Salawu et al.[11], one of the largest english-language abusive tweet corpuses online. We recognize that by using this dataset there are limitations in our training, for example, of more subtle abuse not involving the key words used to label tweets (including the Google Swear Words List as originally used by Salawu et al.). Additionally, a limitation of this research more generally is that it cannot, by the nature of social media messaging, capture private messages. It is known from the pair qualitative research that much abuse exists in the private messaging domain.

The associated tweets for each ID are then pulled directly from the twitter API V2 using the get_tweet method from the Tweepy package in Python.

### Pre-Processing

Having acquired 'abusive' tweets using this dataset, contextual information is needed to enrich the user profiles using them. 'Interaction pairs' of abusers and targets are extracted, and additional contextual information is generated from each tweet's associated user profiles. Approximated labels for genders are applied, and female-only target tweets and interactions involving at least one female (ruling out male-male interactions) are extracted as below.

### Gender Labelling: Abuser and Originator

The first step is to provide gender labels as Twitter does not explicitly record users' gender information. This is thus achieved through running each users first name through the Gender API (https://gender-api.com/), as suggested by the methodology presented in Salawu et al.[12]. This is assumed to provide a reasonable and functional approximation of gender identity for a given time. A further limitation is that this can only account for binary gender classification- we cannot currently extract non-binary gender types.

### Extraction of Target

Having now achieved a dataset of gender-labelled tweets, a second iteration is performed to look for the 'target'; i.e. other accounts mentioned within a single tweet. This is achieved through performing a regular expression search for '@' and then the subsequent string until a space is found. Any further mentions are separated into the originator, target 1, target 2, down to target 5- as no more than 5 mentions are found in any of the tweets in the dataset.

The gender labelling is then repeated on each of the target accounts.

**Multidimensional Indicator Extraction**

Having achieved a gender and directionality labelled dataset, interaction pairs can be explored through the following indicators in order to train an algorithm to detect and recommend removal or action:

- **Longitudinal**: the number of times that a prior abusive interaction occurred between the same abuser and target(s), the direction of the interaction (i.e. originator to target, or target to originator)
- **Informational Asymmetry**: examine associated profile biographic information, extract number of characters in profile biography, number of links present, and other strings associated with public information, for each interaction pair - label the % information shared on the target vs. the abuser feeds
- **Volumetric asymmetry**: examine the number of tweets in each direction (from originator to target, and target to originator), and calculate ratio as a % of directionality of tweets, extract number of tweets directed to the target in a user-specified elapsed time period.

**Detection Algorithm**

The proposed moderation algorithm ingests a normal tweet and examines the content for the above indicators. The 'user' is assumed to be the target, who is using the moderation algorithm to detect behaviors associated with these abusive indicators. The following text prompts and interventions are then suggested for user input:

- Longitudinal abuse - *'This person has tweeted you X times before- would you like to block them?'*
- Informational Asymmetry - *'This account has very little information on it- would you like to block them?'*
- Volumetric asymmetry - *'You are receiving an unusual volume of tweets for your profile. Would you like to delete all incoming tweets?'*

**Perspective API for content moderation -** use of the publicly available Perspective API for detection of common toxicity and abusive behaviors, to provide a combined immediate and multi-dimensional abuse detection.

**IV. PROPOSED PIPELINE**

Fig. A outlines the resulting technical architecture in detail.¶

FIG A. PROPOSED END-TO-END DECENTRALISED MODERATION ALGORITHM ARCHITECTURE FOR FEMALE-TARGETED ONLINE ABUSE

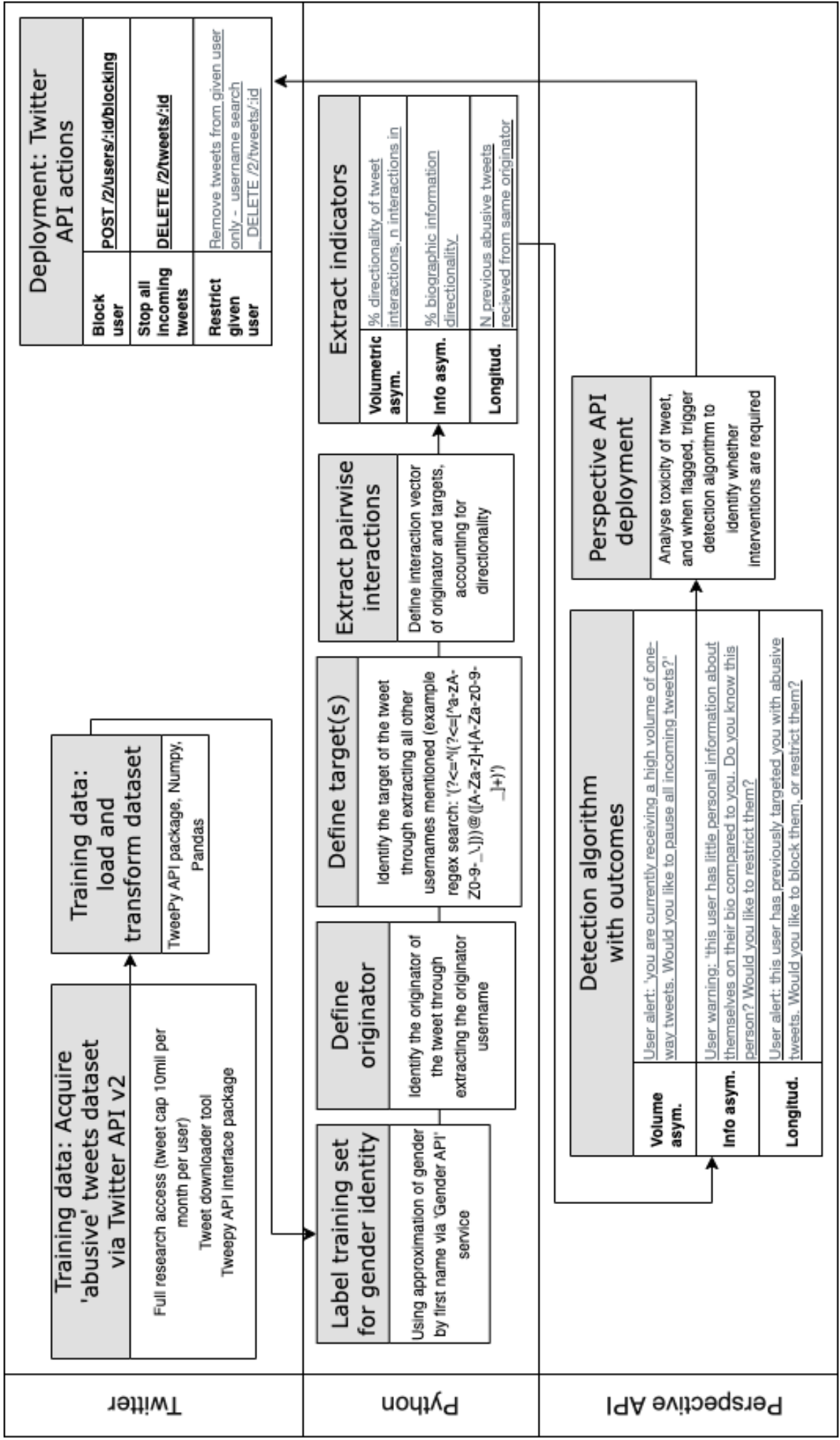

## V. FURTHER WORK

This paper proposes a work in progress algorithm, with the architecture in Fig A. being the main product of this work. In the next phase, platform and tweet snapshots will be added to bring life to the multidimensional indicators listed above, alongside testing specific API endpoints to work in unison. Additionally, further discussion will be added to further outline the details of each of the limitations identified in the above article, and suggest further work to address these.

## VI. ACKNOWLEDGEMENTS

The authors extend thanks to the Center for Long Term Cybersecurity, UC Berkeley, for the financial and academic support that made this work possible.